\documentclass[lettersize,journal]{IEEEtran}

\RequirePackage{conor} 

\title{Libfork: portable continuation-stealing with \\ stackless coroutines}

\author{C.J. Williams, J.A. Elliott}

\begin{document}

\maketitle

\begin{abstract}
Fully-strict fork-join parallelism is a powerful model for shared-memory programming due to its optimal time-scaling and strong bounds on memory scaling. The latter is rarely achieved due to the difficulty of implementing continuation-stealing in traditional High Performance Computing (HPC) languages -- where it is often impossible without modifying the compiler or resorting to non-portable techniques. We demonstrate how stackless-coroutines (a new feature in \Cppn{\bm{20}}) can enable fully-portable continuation stealing and present \emph{libfork} a lock-free fine-grained parallelism library, combining coroutines with user-space, geometric segmented-stacks. We show our approach is able to achieve optimal time/memory scaling, both theoretically and empirically, across a variety of  benchmarks. Compared to openMP (libomp), libfork is on average $7.2\times$ faster and consumes $10\times$ less memory. Similarly, compared to Intel's TBB, libfork is on average $2.7\times$ faster and consumes $6.2\times$ less memory. Additionally, we introduce non-uniform memory access (NUMA) optimizations for schedulers that demonstrate performance matching \emph{busy-waiting} schedulers.
\end{abstract}

\section{\label{sec::intro}Introduction}

\IEEEPARstart{S}{hrinking} transistors are the historic driving-force behind the rapid increase in computing power. Moore's law forecasts an exponential growth in the number of transistors on an integrated circuit~\cite{Moore2006}; as we reach the physical limits of transistor size, further increases in compute power are achieved by increasing the number of logical cores on a single chip~\cite{Kish2002}. Programs must embrace parallelism to take advantage of this horizontal scaling.

Parallel computers commonly come in two flavours: shared-memory computers have a single address-space shared by all cores; distributed-memory computers have separate address spaces for each core (normally these are physically separated). Different programming paradigms are appropriate for each type of computer. In this paper we focus on shared-memory parallelism (SMP) which represents almost all modern multi-core computers. Furthermore, distributed-memory computers usually employ two-level parallelism, leveraging SMP within compute nodes and falling back to message-passing between nodes. Therefore, SMP is relevant even for many distributed-memory HPC systems.

Complexities, such as instruction/memory (re)ordering, cache coherency and synchronization overhead, make low-level SMP programming difficult to reason about and error prone. Hence, higher level abstractions -- \ie structured concurrency -- have been developed to insulate the programmer from these complexities. Examples include: data parallelism~\cite{Hillis1986}, pipeline parallelism~\cite{Lee2015}, task parallelism~\cite{Thoman2018}, the actor model~\cite{Actor1987}, \etc. Choosing a parallel programming model has a large impact on the performance and
complexity of a program.

A parallelism framework expresses the available concurrency of a program, when and where synchronization is needed and, sometimes where code must be executed. It is the job of a scheduler to execute the program in parallel on the physical hardware. If the problem or hardware allow it, the execution can be scheduled offline/statically or even at compile-time, otherwise it must be scheduled online/dynamically at runtime. Scheduling on heterogenous computers almost always requires online scheduling, as does irregular/unpredictable workloads. Therefore, we focus on online scheduling as the more general paradigm.

The remainder of this paper is structured as follows: \cref{sec::back} provides the necessary background; \cref{sec::libfork} details our \Cpp library, libfork, and how the operations of continuation stealing can be efficiently mapped to stackless coroutines; in \cref{sec::eval} we evaluate the performance of libfork; finally, in \cref{sec::conc} we draw conclusions and discuss future work. 
\section{\label{sec::back}Background}

The background section is divided into four components: in \cref{sec::coro} we introduce generic coroutines and  \Cppn{20}'s  stackless variation; \cref{sec::fj} introduces the fork-join model of parallelism; in \cref{sec::ws} we expand upon work-stealing and its associated data-structures finally, in \cref{sec::cac} we introduce cactus stacks and the challenges they present.

\subsection{\label{sec::coro}Coroutines}

Semantically, a coroutine is a function with the ability to pause execution (with the \texttt{suspend} operation), which may then be continued at a later time (with the \texttt{resume} operation). Coroutines are commonly used in asynchronous programming and cooperative (multi)tasking. They are available in many higher-level programming languages such as Python~\cite{Sanner1999PythonAP}, Golang~\cite{Pike2012}, Kotlin~\cite{Elizarov2021}, \etc.

A full taxonomy of coroutines is presented by \citeauthor{Moura2009}~\cite{Moura2009}. In brief, coroutines can be either stackless or stackfull. A stackfull coroutine (also called a green thread or fibre) is a user space equivalent of an operating system (OS) thread, similarly equipped with its own stack. As stackfull coroutines are scheduled cooperatively, the cost of a context switch is orders of magnitude faster than an OS thread. In contrast, stackless coroutines do not have their own stack. Instead, variables that span a suspension point are stored in a \emph{coroutine frame}. This parallels (but does not replace) the stack frame of a regular function. This means a stackless coroutine cannot \texttt{suspend} from within a nested (regular) function call. In general, a coroutine frame must be dynamically allocated/deallocated. The context switch between stackless coroutines can be almost as fast as a bare function call.

\subsubsection{Coroutines in \Cpp}

\Cppn{20} introduces stackless coroutines (with both symmetric and asymmetric control transfer mechanisms) however, they expose their \texttt{suspend}/\texttt{resume} operations indirectly. For the purpose of exposition, a \Cppn{20} coroutine:
\begin{itemize}
    \item Is a function which allocates a (compiler generated) coroutine frame (this allocation can be overridden by the user) at the point of invocation.
    \item Can \mintinline{cpp}{co_await} an \emph{awaitable} object. Effectively, this calls \texttt{suspend} on the current coroutine's frame and passes the suspended frame to the awaitable. The awaitable is then free to transfer control to another coroutine (by calling \texttt{resume} on a frame) or \mintinline{cpp}{return} to the previous \texttt{resume} call.
    \item Terminates with a \mintinline{cpp}{co_return} statement which: optionally returns a value via its coroutine frame and then  \mintinline{cpp}{co_await}(s) a user-defined final-awaitable.
\end{itemize}
Transferring control from within a suspended coroutine to another suspended coroutine (via a \texttt{resume} operation) can use \emph{symmetric-transfer}, this is a guaranteed tail-call, thus consuming no OS stack space.

\subsection{\label{sec::fj}The fork-join model of parallelism}

The fork-join (FJ) model~\cite{Conway1963, Nyman2016} is a popular form of structured task-based concurrency used in programming languages and libraries such as: cilk~\cite{Frigo1998}, openMP~\cite{Ayguade2009}, taskflow~\cite{Huang2019}, Intel's TBB~\cite{Kukanov2007}, nowa~\cite{Schmaus2021}, fibril~\cite{Yang2016} and many others.

Within the task-based concurrency paradigm, a task is an abstract unit of work that can be executed concurrently with other tasks. A task may have dependencies on other tasks. The FJ model augments a language with the keywords \texttt{fork}\footnote{Not to confused with the Unix \texttt{fork(2)} system call.} (sometime called \texttt{spawn}) and \texttt{join} (sometime called \texttt{sync}). The \texttt{fork} keyword creates a new (child) task which may be executed concurrently with the parent task. The \texttt{join} keyword signals that all child tasks (children) must be completed before execution of the parent task is allowed to continue. The fully-strict FJ model (SFJ) further imposes that all children are complete before the parent task returns/completes~\cite{halpern2012strict}. This constraint appears restrictive, but enables simpler reasoning about the program and stronger bounds on time/memory scaling.

\begin{algorithm}
    \begin{algorithmic}[1]
\Function{Fib}{$n$}
        \If{$n < 2$}
        \State \Return $n$
        \EndIf
        \State $x \gets \textbf{fork } \Call{Fib}{n-1}$
        \State $y \gets \Call{Fib}{n-2}$
        \State \textbf{join}
        \State \Return $x + y$
        \EndFunction
    \end{algorithmic}
    \caption{\label{alg::fib}Pseudocode for the Fibonacci recurrence, \cref{eq::fib}, with fork-join annotations to support parallel execution.}
\end{algorithm}

The canonical example of the SFJ model is the Fibonacci function, presented in \cref{alg::fib}, which computes the Fibonacci recurrence~\cite{lucas1891}:
\begin{align}
    F_n = \begin{cases}
              0                 & \text{if } n = 0 \\
              1                 & \text{if } n = 1 \\
              F_{n-1} + F_{n-2} & \text{otherwise}
          \end{cases} \label{eq::fib}
\end{align}
The second recursive call in \cref{alg::fib} does not use the \texttt{fork} keyword as it is immediately followed by a \texttt{join}. Hence, the continuation of the parent task would contain no work if a \texttt{fork} was added. Adding a \texttt{fork} here would not invalidate the program but would add unnecessary overhead.

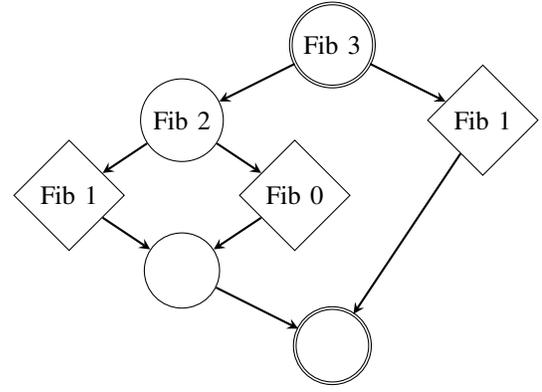
\begin{figure}
    \centering
    \begin{tikzpicture}
        \node[draw, circle, double] (n0) at (0, 0) {Fib $3$};

        \node[draw, circle] (n1) at (-2, -1) {Fib $2$};
        \node[draw, diamond] (n2) at (2, -1) {Fib $1$};

        \node[draw, diamond] (n3) at (-3.5, -2) {Fib $1$};
        \node[draw, diamond] (n4) at (-0.5, -2) {Fib $0$};

        \draw[-stealth, thick] (n0) -- (n1);
        \draw[-stealth, thick] (n0) -- (n2);

        \draw[-stealth, thick] (n1) -- (n3);
        \draw[-stealth, thick] (n1) -- (n4);

        \node[draw, circle, minimum size = 1cm] (f1) at (-2, -3) {};

        \draw[-stealth, thick] (n3) -- (f1);
        \draw[-stealth, thick] (n4) -- (f1);

        \node[draw, circle, double, minimum size = 1cm] (f0) at (0, -4) {};

        \draw[-stealth, thick] (f1) -- (f0);
        \draw[-stealth, thick] (n2) -- (f0);

    \end{tikzpicture}
    \caption{\label{fig::fib_dag} The DAG representing the execution of the Fibonacci function from \cref{alg::fib} with argument $n=3$. Diamond shaped nodes represent leaf tasks/function-calls while circular nodes represent non-leaf tasks, each with a matched join node. Edges represent dependencies \ie parent-child relationships.}
\end{figure}

Each SFJ program maps to a directed acyclic graph (DAG) expressing the dependencies between parents and children. This model is useful for reasoning about the time and memory bounds of a program and is a common intermediate representation for schedulers. The DAG for \cref{alg::fib} is shown in \cref{fig::fib_dag}. A scheduler is free to execute any tasks whose dependencies have been satisfied. Normally, the SFJ model is paired with a work stealing-scheduler (see \cref{sec::ws}) which makes the decision when and where to execute tasks dynamically, as the DAG is built.

The \emph{serial projection} of a SFJ program is a sequential program -- generated by removing the \texttt{fork}/\texttt{join} keywords -- which executes the tasks in a depth first traversal of the DAG. The execution time and (stack) memory-consumption of the serial projection are denoted $T_s$ and $M_s$ respectively.

\subsection{\label{sec::ws}Work stealing schedulers}

A work stealing scheduler (WSS) maps a program's DAG to a set of cores on a physical machine. Each thread of execution (normally each thread is bound to a physical computation core), called a \emph{worker}, is equipped with a work stealing queue (WSQ) (see \cref{sec:wsq} for details). A single worker begins the execution of a program with the \emph{root} task. When a worker encounters a \texttt{fork} statement it creates a new task and pushes a task onto its own WSQ. Other workers can steal from this queue (or any other non-empty WSQ) if they have no tasks to execute. A greedy WSS with $P$ workers can execute a SFJ program in expected time~\cite{Blumofe_1999}:
\begin{align}
    T_p \le \frac{T_1}{P} + \bigO{T_\infty} \label{eq::time_ideal}
\end{align}
where $T_1$ is the time taken to execute the program with a single worker and $T_\infty$ is the ideal runtime on a machine with an infinite number of workers. This expected time is optimal within a constant factor~\cite{Blumofe_1999}. In the DAG model, $T_\infty$  corresponds to the longest path through the DAG. An ideal WSS will use stack space~\cite{Blumofe_1999}:
\begin{align}
    M_p \le PM_1 \label{eq::mem_ideal}
\end{align}
where $M_1$ is the stack space used by a single worker executing the program. This bound can be derived from the DAG, see \citeauthor{Blumofe_1999}~\cite{Blumofe_1999} for full details.

\subsubsection{\label{sec:wsq}Work stealing queues}

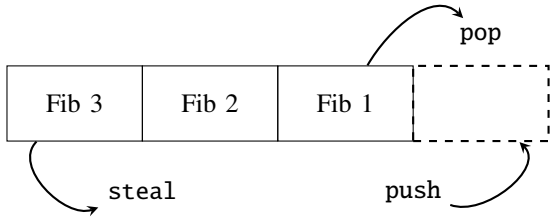
\begin{figure}[tb]
    \centering
    \begin{tikzpicture}[scale=0.9]

        \node[draw, rectangle, minimum width = 1.8cm, minimum height = 1cm] (n3) at (0, 0) {Fib $3$};
        \node[draw, rectangle, minimum width = 1.8cm, minimum height = 1cm] (n2) at (2, 0) {Fib $2$};
        \node[draw, rectangle, minimum width = 1.8cm, minimum height = 1cm] (n1) at (4, 0) {Fib $1$};
        \node[draw, thick, rectangle, dashed, minimum width = 1.8cm, minimum height = 1cm] (n0) at (6, 0) {};

        \node (pop) at (6, 1) {\texttt{pop}};
        \node (push) at (5, -1.3) {\texttt{push}};
        \node (steal) at (1, -1.3) {\texttt{steal}};

        \draw[-stealth, thick] (n1) to  [out=60,in=135] (pop);
        \draw[-stealth, thick] (push) to  [out=-20,in=-45] (n0);
        \draw[-stealth, thick] (n3) to  [out=-135,in=-160] (steal);

    \end{tikzpicture}
    \caption{\label{fig::wsq} A diagram of a work stealing queue, the queue contains handles to tasks during a moment of a depth first traversal of the DAG in \cref{fig::fib_dag}.}
\end{figure}

A work stealing queue is a semi-concurrent data structure, which efficiently supports the operations: \texttt{push}, \texttt{pop} and, \texttt{steal}. The \texttt{push} and \texttt{pop} operations, performed by a single owning worker/thread, add/remove elements of the queue in a first-in-last-out (FILO) order. The \texttt{steal} operation can be performed by any worker and removes an element from the queue in a first-in-first-out (FIFO) order. \Cref{fig::wsq} contains a schematic of a WSQ, the \texttt{steal} operation may be called concurrently with \texttt{push}, \texttt{pop} or other \texttt{steal} operations.

An efficient WSQ is the backbone of a work stealing runtime; the minimum overhead of a task (compared to a bare function call) corresponds to pushing then popping a task (or task handle) to and from a WSQ. Early implementations of WSQs include the THE~\cite{Frigo1998} queue and the ABP~\cite{Arora_2001} queue. Some papers have explored variations of WSQs that: enable stealing half the tasks a worker owns~\cite{Hendler2002}, have split public and private sections~\cite{vanDijk2014, Dinan2009, Cartier2021} and, can store the tasks inline as opposed to pointers/handles to them~\cite{Faxn2008, Faxen2010}. These variations all have their merits however, we use a modern version~\cite{L2013, Norris_2013} of the Chase-Lev (CL)~\cite{Chase2005} queue which is fully lock-free, optimized for weak memory models and formally verified~\cite{L2013, Choi2023FormalVO}.

\subsubsection{Child vs continuation stealing}

When a worker forks a task it can choose either to execute the child task and push the \emph{continuation} of the parent task onto it's WSQ or it can push the child task onto its WSQ and continue execution of the parent. The former corresponds to a depth first traversal of the DAG and is called \emph{continuation stealing}, while the latter corresponds to a breadth first traversal and is called \emph{child stealing}. Child stealing is more common, as it is possible to implement as a library in most programming languages~\cite{Shiina2022, Schmaus2021}. However, child stealing breaks the memory bound, \cref{eq::mem_ideal}, as a task can have an unbounded number of children and each child will require some memory. Therefore, continuation stealing is the preferable strategy. Continuation stealing can also lead to better cache locality as the child task is likely to use data that the parent task has loaded into memory. Furthermore, continuation stealing preserves the order of execution between a programs's serial projection and its single-worker execution.

\subsection{\label{sec::cac}Cactus stacks}

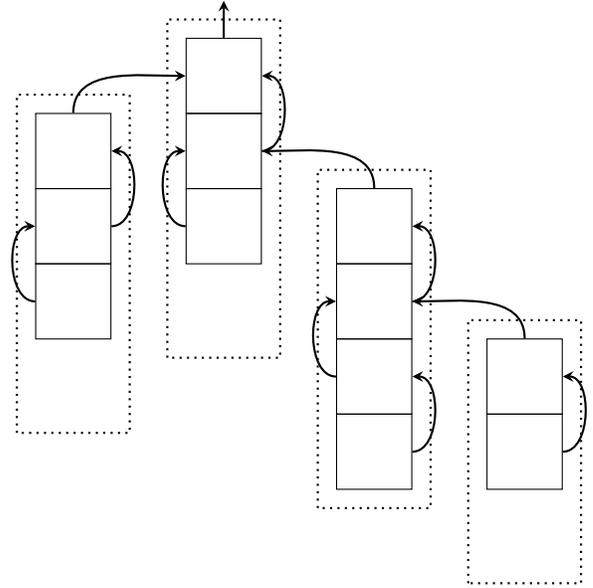
\begin{figure}[tb]
    \centering
    \begin{tikzpicture}

        \node[draw, rectangle, minimum width = 1cm, minimum height = 1cm] (a0) at (2, 1) {};
        \node[draw, rectangle, minimum width = 1cm, minimum height = 1cm] (a1) at (2, 2) {};
        \node[draw, rectangle, minimum width = 1cm, minimum height = 1cm] (a2) at (2, 3) {};

        \draw[dotted, thick] (1.25, -0.75) rectangle (2.75, 3.75);

        \draw[-stealth, thick] (a0) to [out=180,in=180] (a1);
        \draw[-stealth, thick] (a1) to [out=0,in=0] (a2);

        \draw[-stealth, thick] (a2) -- (2, 4);

        \node[draw, rectangle, minimum width = 1cm, minimum height = 1cm] (n0) at (0, 0) {};
        \node[draw, rectangle, minimum width = 1cm, minimum height = 1cm] (n1) at (0, 1) {};
        \node[draw, rectangle, minimum width = 1cm, minimum height = 1cm] (n2) at (0, 2) {};

        \draw[dotted, thick] (-0.75, -1.75) rectangle (0.75, 2.75);

        \draw[-stealth, thick] (n0) to [out=180,in=180] (n1);
        \draw[-stealth, thick] (n1) to [out=0,in=0] (n2);
        \draw[-stealth, thick] (n2) to [out=90,in=180] (a2);

        \node[draw, rectangle, minimum width = 1cm, minimum height = 1cm] (b0) at (4, -2) {};
        \node[draw, rectangle, minimum width = 1cm, minimum height = 1cm] (b1) at (4, -1) {};
        \node[draw, rectangle, minimum width = 1cm, minimum height = 1cm] (b2) at (4, 0) {};
        \node[draw, rectangle, minimum width = 1cm, minimum height = 1cm] (b3) at (4, 1) {};

        \draw[-stealth, thick] (b3) to [out=90,in=0] (a1);
        \draw[-stealth, thick] (b2) to [out=0,in=0] (b3);
        \draw[-stealth, thick] (b1) to [out=180,in=180] (b2);
        \draw[-stealth, thick] (b0) to [out=0,in=0] (b1);

        \draw[dotted, thick] (3.25, -2.75) rectangle (4.75, 1.75);

        \node[draw, rectangle, minimum width = 1cm, minimum height = 1cm] (d0) at (6, -1) {};
        \node[draw, rectangle, minimum width = 1cm, minimum height = 1cm] (d1) at (6, -2) {};

        \draw[-stealth, thick] (d0) to [out=90,in=0] (b2);
        \draw[-stealth, thick] (d1) to [out=0,in=0] (d0);

        \draw[dotted, thick] (5.25, -3.75) rectangle (6.75, -0.25);

    \end{tikzpicture}
    \caption{\label{fig::cactus} A diagram of a cactus stack, sometimes called a spaghetti stack, which is an example of a parent pointer tree. Boxes represent (stack) frames. Arrows denote a parent-child relationship. The dotted lines represent regions that could be contiguous segments of memory, \ie the first child of each parent can be placed on the parents (linear) stack.}
\end{figure}

In the context of the SFJ model, each task has an associated \emph{frame} which contains variables local to the task. These are equivalent to traditional stack-frames in the serial projection of a program. A frame may contain pointers into itself hence, it is sensitive to its own location in memory and cannot generally be relocated. A given task may contain pointers to variables in any frame \emph{above} it (\ie its parent, grandparent, \etc) just like a traditional stack frame. As we are restricted to SFJ the lifetime of a child frame must strictly nest the lifetime of its parent.

For a continuation stealing worker, when no steals are occurring, each frame can be mapped onto a linear stack -- as in the serial projection. However, after a worker steals a task, it must allocate the frame of any newly forked children (\ie not the first child) on a new stack, this produces the branches in \cref{fig::cactus}. The parent's stack cannot be used by the thief as it may still be in use by another worker, executing some other descendent of the parent. The new stack is linked to the parent stack. The resulting tree-of-stacks data structure is called a cactus stack~\cite{Clinger1988} and is sketched in \cref{fig::cactus}. Designing a cactus stack that is simultaneously:
\begin{enumerate}
    \item \label{item::interop} Interoperable with a legacy/linear stack
    \item Supports a linear-scaling scheduler
    \item Provides bounded and efficient memory use
\end{enumerate}
is an open research problem~\cite{Lee2010, Yang2016}. In \cref{sec::stack} we present a partial-solution to this problem that forgoes \cref{item::interop} in favour of performance and compatibility with stackless coroutines.

\section{\label{sec::libfork}Libfork}

Libfork is pure \Cppn{20} library that implements lock-free, wait-free, continuation-stealing using stackless coroutines. The application programmer interface (API) of libfork is designed to mirror Cilk~\cite{Frigo1998} and be as unsurprising as possible. The core API is summarized in \cref{alg::fib_lf}, which presents the fibonacci function from \cref{alg::fib} expressed with libfork's primitives.

\begin{algorithm}
    \begin{minted}[
autogobble
    ]{cpp}
    auto fib = []
      (auto fib, int n) -> task<int> {
       
        if (n < 2) {
            co_return n;
        }
        
        int a, b;

        co_await fork[&a, fib](n - 1);
        co_await call[&b, fib](n - 2);

        co_await join;

        co_return a + b;
      };
    \end{minted}
    \caption{\label{alg::fib_lf} The fibonacci function from \cref{alg::fib} written in \Cpp with the libfork library. Note: namespace qualifiers and optional decorators have been omitted for brevity. The first argument to a coroutine passes static information through the type-system (\eg if the task is a root task, the invocation kind, the type of its return address, \etc) and acts as a y-combinator, allowing for recursion within anonymous functions.}
\end{algorithm}

Libfork differs from all other \Cpp/C libraries of its kind because it is a fully-portable, library-only implementation with strong bounds on time and memory scaling, see \cref{sec::theory}. Libfork utilizes segmented stacks to store tasks (detailed in \cref{sec::stack}) which allow for unbounded task recursion, without the fear of stack overflows. Additionally, libfork's stacks are exposed to the user to allow for portable use of a (safer) \texttt{alloca(3)} equivalent.

Within libfork, a stackless coroutine corresponds to a task. The mapping between the operations of continuation stealing fork-join parallelism and the operations of stackless coroutines in combination with user-space stacks are detailed in section \cref{sec::map}.

Finally, libfork is fully NUMA aware and introduces a variation of the adaptive scheduler (see \cref{sec::numa}) presented by \citeauthor{Lin2020}~\cite{Lin2020} which is able to match the performance of busy-waiting schedulers on NUMA machines.

\subsection{\label{sec::stack}Segmented stacks}

As libfork uses continuation stealing, each worker is either a \emph{thief} with no tasks or \emph{active} and there exists a chain of tasks from the root task to the worker's currently executing task. The coroutine frames along this chain, called a \emph{strand}, are linked into a cactus stack.

\Cref{fig::cactus} gives an example of a cactus stack; the most straightforward implementation allocates heap memory for every coroutine frame. This upholds the memory bound and allocations/deallocations can be $\bigO{1}$ however, heap allocations are comparatively costly operations. Perhaps worse, the strand could be fragmented in memory, leading to inferior cache-locality compared to a traditional linear stack.

Another potential solution is to use a large linear stack, placing the first child of each parent on the parents linear stack and allocating new linear stacks as-required at branch points. This enables the fastest possible allocations/deallocations and the best possible cache locality along the fast-path. The size of the linear stacks would have to be $\bigO{M_1}$, in-case no branching occurred. Unfortunately, in the worst case, almost every coroutine frame could be placed on a different linear-stack. The memory bound would therefore approach:
\begin{align}
    M_p \le P M_1^2
\end{align}
As $M_1$ is $\bigO{10^6}$ this is much worse than the bound discussed in \cref{sec::ws}. The memory bound would be restored if the wasted space in the linear stacks could be reclaimed. This is possible with segmented stacks~\cite{Ma2023} which grow/shrink on demand.

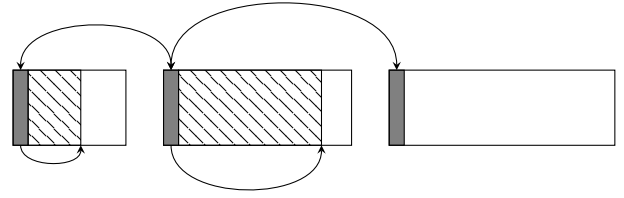
\begin{figure}[tb]
    \centering
    \begin{tikzpicture}

        \draw[draw=black, fill=gray] (0, 0) rectangle (0.2, 1);
        \draw[draw=black, fill=gray] (2, 0) rectangle (2.2, 1);
        \draw[draw=black, fill=gray] (5, 0) rectangle (5.2, 1);

        \draw[draw=black, pattern={Lines[angle=-45,distance=4pt]}] (0.2, 0) rectangle (0.9, 1);
        \draw[draw=black, pattern={Lines[angle=-45,distance=4pt]}] (2.2, 0) rectangle (4.1, 1);

        \draw[draw=black] (0, 0) rectangle (1.5, 1);
        \draw[draw=black] (2, 0) rectangle (4.5, 1);
        \draw[draw=black] (5, 0) rectangle (8, 1);

        \draw[stealth-stealth] (0.1, 1) to [out=90,in=90]  (2.1, 1);
        \draw[stealth-stealth] (2.1, 1) to [out=90,in=90]  (5.1, 1);

        \draw[-stealth] (0.1, 0) to [out=-90,in=-90]  (0.9, 0);
        \draw[-stealth] (2.1, 0) to [out=-90,in=-90]  (4.1, 0);

    \end{tikzpicture}
    \caption{\label{fig::stack} Diagram (not to scale) of a segmented stack in libfork. The metadata region is filled in gray, hatched regions indicate allocated space, double ended arrows indicate doubly-linked list connections and, single ended arrows represent each stacklets' stack-pointer. This stack is composed of three stacklets, the middle stacklet is the top stacklet, \ie contains the last allocation. The rightmost stacklet is a cached stacklet, each stack contains zero-or-one cached stacklets.}
\end{figure}

Libfork's utilizes geometric segmented-stacks. A stack is composed of segments of contiguous memory called \emph{stacklets}. Each stacklet starts with \num{48}\si{\byte} of metadata containing pointers: linking the stacklets into a doubly-linked list, tracking the position of the stacklets internal stack-pointer and, marking the region of memory available to the stacklet. If an allocation can fit on the current/top stacklet then an allocation is as fast as a pointer increment. Otherwise, a new stacklet, twice as large as the previous one (or large enough to fit the allocation, whichever is greater), is allocated from the heap. The time for $n$ consecutive allocations is:
\begin{align}
    n T_\text{pointer} + \bigO{\log_2{n}} T_\text{heap}
\end{align}
hence, the amortized cost of a single allocation is $\bigO{T_\text{pointer}}$. If a stacklet becomes empty after a deallocation then it may be cached (if it is not more than twice as large as the previous stacklet). This guards against hot-splitting~\cite{Ma2023}. The allocation and deallocation hot paths are identical to a linear stack, the additional instruction cost is loading more pointers and a predictable branch.

\subsection{\label{sec::map}Continuation stealing with stackless coroutines}

\subsubsection{Fork or call}

The call to the \texttt{fork} function in \cref{alg::fib_lf} creates a new child task, which will recursively call \texttt{fib} with a new argument and write the results to the variable $a$ in the parent task. The \texttt{fork} function generates an awaitable who's actions are specified in \cref{alg::fork}.
\begin{algorithm}
    \begin{algorithmic}[1]
\Require $p \gets$ the parent's suspended frame.
        \Require $f \gets$ a child function.
        \Require $x\ldots \gets$ arguments for $f$.

        \Function{AwaitFork}{$p$, $f$, $x\ldots$}

        \State $g \gets$ the \textbf{thread-local} stack
        \State $s \gets$ allocate space for child on $g$
        \State $c \gets$ invoke $f$ with $x\ldots$ (frame at $s$ on $g$) \Comment{Child}

        \State Set $p$ as the parent of $c$
        \State Set $g$ as the stack of $c$

        \State Push $p$ onto the \textbf{thread-local} WSQ \label{line::push}

        \State \textbf{tailcall} \Call{Resume}{$c$} \Comment{Execute child}

        \EndFunction
    \end{algorithmic}
    \caption{\label{alg::fork}Pseudocode for the fork-awaitable, return address handling omitted for brevity.}
\end{algorithm}
The thread-local objects (the stack and WSQ) are stored in \mintinline{cpp}{thread_local} variables, such that they can be accessed without returning control to the scheduler. Only the push to the WSQ on \cref{line::push} requires atomic/synchronized operations. The call-awaitable, generated by the \texttt{call} function in \cref{alg::fib_lf}, is identical to \cref{alg::fork} except the push to the WSQ is omitted. The tail-call is achieved through symmetric-transfer as the child, $c$, is constructed in the suspended state. In general, the child task \emph{may} be allocated on a different frame than the parent.

\subsubsection{Join}

\begin{algorithm}
    \begin{algorithmic}[1]
\Require $c \gets$ a suspended coroutine frame.
        \Function{AwaitJoin}{$c$}

        \If{this task has not been stolen}
        \State \textbf{tailcall} \Call{Resume}{$c$} \Comment{already own $c$'s stack}
        \EndIf

        \Atomic \label{line::atomic_join}
        \State Mark $c$ as at-a-join-point
        \State Decrease $c$'s join counter
        \EndAtomic

        \If{this was the last worker to join $c$}
        \State $g \gets$ the \textbf{thread-local} stack
        \State $g_c \gets$ the stack of $c$
        \State $g \gets g_c$ \label{line::take_join} \Comment{take $c$'s stack}
        \State \textbf{tailcall} \Call{Resume}{$c$}
        \EndIf

        \State \textbf{return} \Comment{to scheduler, }

        \EndFunction
    \end{algorithmic}
    \caption{\label{alg::join}Pseudocode for the join-awaitable.}
\end{algorithm}

The join-awaitables's actions are described in \cref{alg::join}. We use the split counter method of nowa~\cite{Schmaus2021} to implement the operations in the atomic block, \cref{line::atomic_join}, with no locks. If the worker is the last to join, it takes ownership of the child's stack on \cref{line::take_join}; before taking ownership the workers previous stack, $g$, is empty. We implement an additional shortcut/optimization before entering \cref{alg::join} to prevent unnecessary suspension of the parent task when no steals have occurred or all children are already complete. All tail-calls are achieved through symmetric-transfer.

\subsubsection{Cooperatively returning}

\begin{algorithm}
    \begin{algorithmic}[1]
\Require $c \gets$ a suspended coroutine frame.
        \Function{AwaitReturn}{$c$}

        \State $p \gets$ the parent of $c$
        \State $g \gets$ the \textbf{thread-local} stack
        \State $g_p \gets$ the stack of $p$

        \State Deallocate $c$ from $g$

        \If{$p$ is null} \Comment{\ie $c$ is a root task}
        \State \textbf{return} \Comment{to scheduler}
        \EndIf

        \If{$c$ was called}
        \State \textbf{tailcall} \Call{Resume}{$p$}
        \EndIf

        \If{try \texttt{pop} from thread local WSQ}
        \State \textbf{tailcall} \Call{Resume}{$p$} \Comment{hot-path}
        \EndIf

        \Atomic \Comment{implicit join}
        \State Check if $p$ is at-a-join-point
        \State Decrease $p$'s join counter
        \EndAtomic

        \If{$p$ was at-a-join-point}
        \If{this was the last worker to join $p$}
        \If{$g \ne g_p$}
        \State $g \gets g_p$ \label{line::take_final}  \Comment{take $p$'s stack}
        \EndIf
        \State \textbf{tailcall} \Call{Resume}{$p$}
        \EndIf
        \EndIf

        \If{$g = g_p$}
        \State $g \gets$ new empty stack \label{line::release}  \Comment{release $p$'s stack}
        \EndIf

        \State \textbf{return} \Comment{to scheduler}

        \EndFunction
    \end{algorithmic}
    \caption{\label{alg::final}Pseudocode for the final-awaitable, return-value handling omitted for brevity.}
\end{algorithm}

The final-awaitables's actions are described in \cref{alg::join}. The first and second \texttt{if}-statements can be resolved at compile time (using the static information discussed in \cref{alg::fib_lf}) and thus have no runtime-overhead. Again, the split counter method~\cite{Schmaus2021} is used to implement the atomic block with no locks. If the worker fails to pop the parent from its WSQ, it must perform an \emph{implicit join}, potentially transferring execution back to the parent. If the implicit-join succeeds and the worker does not already own the parents stack, the worker takes ownership of the parents stack on \cref{line::take_final}. Before taking ownership, the workers previous stack, $g$, is empty. Otherwise, if the implicit-join fails, the worker may need a new stack. If the worker does need a new stack, the old one is released on \cref{line::release} (some worker will eventually take ownership of it when resuming the parent). Once again, all tail-calls are achieved through symmetric-transfer.

\subsection{\label{sec::stalloc}Stack allocation API}

A fork-join scope is the region between the first \texttt{fork} and its corresponding \texttt{join}. Outside a fork-join scope, a worker always "owns" the stack a coroutine lives on. Hence, they can allocate/deallocate from it -- as long as they preserve FILO ordering and strictly nest the lifetime of the allocations within the coroutine's lifetime. This is useful for allocating temporary memory for storing the return values used in a fork-join scope, \eg a buffer for a parallel reductions partial sums. This is a portable alternative to \texttt{alloca(3)}, which is not available on all platforms and, due to the use of segmented stacks, will never overflow the stack.

\subsection{\label{sec::numa}Scheduling and NUMA}

Libfork's workers are NUMA aware. They use hwloc~\cite{Broquedis2010} to determine the NUMA topology of a machine, which is represented as a tree with physical CPU cores at its leaves. The topological distance between two leaves/cores is the maximum of the distances between each leaf and their common ancestor. Cores that are (topologically) closer have faster access to common shared memory. In libfork, each worker is pinned to a core. When a worker attempts to steal a task from a \emph{victim}, the victim is selected with a probability proportional to:
\begin{align}
    w_{ij} = \frac{1}{n_{ij} r^2_{ij}} \label{eq::victim}
\end{align}
where $r_{ij} \in \mathbb{Z}$ is the topological distance between cores $i$ and $j$ and, $n_{ij}$ is the number of cores separated by $r_{ij}$.

Libfork supplies two default schedulers\footnote{Libfork's extension-API supports user customization of schedulers.}, both of which are greedy~\cite{Blumofe_1999}. The \emph{busy} scheduler's workers continuously attempt randomized stealing, selecting their victims according to \cref{eq::victim}. This minimizes latency at the cost of high CPU usage, even when workers have no tasks. The \emph{lazy} scheduler is a simple variation of the adaptive scheduler presented by \citeauthor{Lin2020}~\cite{Lin2020}. We separate workers into groups as determined by the NUMA node they are pinned to. When at-least one worker is active globally then within each group -- as opposed to globally as suggest by  \citeauthor{Lin2020}~\cite{Lin2020} -- at least one worker is kept awake and attempts randomized stealing. The remaining workers in each group can sleep when/if they find no work. All the workers select their victim according to \cref{eq::victim}. This trades potentially-higher latency for lower CPU usage when the available parallelism is low. Keeping one worker awake per NUMA node reduces cross-node stealing.

\subsubsection{\label{sec::explicit}Explicit scheduling}

Libfork's workers each maintain a WSQ and a lock-free single-consumer multi-producer \emph{submission queue}. This means there is no global submission-queue -- which is often employed to submit root-tasks to the pool of workers. As each task in libfork is a coroutine, workers can use these submission queues to perform \emph{explicit-scheduling}; if a task requests to be run on a particular worker then it can be suspended and ownership transferred by pushing a handle onto the requested workers submission queue. This is desirable for certain runtimes (\eg MPI~\cite{walker1996mpi}) which may require a specific thread to interact with them.

\subsection{\label{sec::theory}Theoretical bounds}

\begin{definition}
    A stacklet has a stack of size $k$ and a metadata region of size $c$. All "sizes" are in bytes unless specified otherwise.
\end{definition}

\begin{theorem}[Segmented stack overhead]
    \label{thm::stack}
    A segmented stack storing $M \ge 1$ bytes has a worst-case size of $\bigO{c} + c\log_2{\left(M\right)} + 4M$.
\end{theorem}

\begin{proof}
    Each allocation must be at-least one byte. There exists $n + 1$ stacklets and $n$ can always be made greater than zero by adding a cached (empty) stacklet. The first $n$ stacklets must contain at least one allocation each. The wasted space on each of the first $n$ stacklets is at most one-less than the size of the requested (stack) allocation that triggered the (heap) allocation of the next stacklet. Hence, the total wasted stack-space on the first $n$ stacklets is at most $M$. Furthermore, the total used stack-space on the first $n$ stacklets is at most $M$ if the last stack is empty. Therefore, the total stack size of the first $n$ stacklets is at most $2M$.
    Summing over the minimal stack-size of the first $n$ stacklets:
    \begin{align}
        1 + 2 + 4 + \ldots + 2^{n - 1} \le 2M
    \end{align}
    hence, the maximum value of $n$ is:
    \begin{align}
        n \le \lfloor\log_2{\left(2M + 1\right)}\rfloor
    \end{align}
    The last stacklet can be at-most twice as large as the previous stacklet hence, its worst case size is $2M$. Therefore, including the metadata overhead, the total memory used by the segmented stack, $M^\prime$, is at most:
    \begin{align}
        M^\prime & \le (n + 1)c + 2M + 2M \nonumber                  \\
                 & \le + c\log_2{\left(2M + 1\right)} + 4M \nonumber \\
                 & \le \bigO{c} + c\log_2{\left(M\right)} + 4M
    \end{align}
\end{proof}

\begin{lemma}
    \label{lem::sum}
    Given a path through a DAG composed of $n$ tasks each consuming memory $m_i$, the total memory consumed by a strand along that path is at-worst $M \le 2nc + 3(m_1 + \ldots + m_n)$.
\end{lemma}

\begin{proof}
    Tasks along a strand are stored on a chain of segmented stacks. The gradient of the result of \cref{thm::stack} with respect to $M$ is monotonically decreasing hence, the largest memory overhead occurs when each task is stored on it's own stack. Therefore, each task requires at-worst:
    \begin{align}
        (c + m_i) + (c + 2m_i) = 2c + 3m_i
    \end{align}
    bytes of memory for a three stacklet stack. Therefore, the memory consumed by the strand along this path is at-worst:
    \begin{align}
        M_\text{path} & \le (2c + 3m_1) + \ldots + (2c + 3m_n) \nonumber \\
                      & \le 2nc + 3(m_1 + \ldots + m_n)
    \end{align}

\end{proof}

\begin{definition}
    Let $M_1$ be the sum of the task sizes along the path through the DAG that maximizes such a sum, \ie $M_1$ is the maximum (stack) memory usage of a program executed by a single worker on a linear stack with no overhead.
\end{definition}

\begin{lemma}
    \label{lem::longest}
    The longest path through a programs's DAG contains at-most $N \le M_1$ tasks.
\end{lemma}

\begin{proof}
    The smallest task must consume at-least $1$ byte of memory. The longest path must consume less memory than the path that uses $M_1$ memory, hence:
    \begin{align}
\sum_{i=1}^{N} 1 & \le M_1
    \end{align}
\end{proof}

\begin{theorem}[Stack memory bound]
    \label{thm::memory}
    Libfork will use at most $M_p \le \left(2c + 3\right) P M_1$ stack memory executing a program with $P$ workers.
\end{theorem}

\begin{proof}
    Due to symmetric transfer, each worker uses no OS-stack space when transferring control between tasks hence, a constant amount of OS-stack space per worker is used. In the worst-case, the path through the DAG that uses $M_1$ memory is also the longest path in the DAG hence, applying \cref{lem::sum} and \cref{lem::longest}, all paths use less than:
    \begin{align}
        M_\text{max} & \le 2nc + 3(m_1 + \ldots + m_n) \nonumber \\
                     & \le 2Nc + 3M_1 \nonumber                  \\
                     & \le \left(2c + 3\right)M_1
    \end{align}
    bytes of memory. Libfork maintains the \emph{busy-leaves property} \ie ``at every time step, every living [task] that has no living descendants has a processor working on it''~\cite{Blumofe_1999}. Therefore, every worker consumes less memory than the path through the DAG that consumes the most memory. Hence, the worst-case memory usage of a program with $P$ workers is:
    \begin{align}
        M_p & \le P M_\text{max} \nonumber  \\
            & \le \left(2c + 3\right) P M_1
    \end{align}
\end{proof}

Our final memory bound is very loose, the constant multiplier $\left(2c + 3\right)$ is dominated by the $2c$ contribution from the metadata overhead. In practice, this contribution is negligible as the average task size is a few hundred bytes and most stacklets store several tasks.

\section{\label{sec::eval}Experimental Evaluation}

\begin{figure*}
    \centering
    \begin{subfigure}[b]{0.49\textwidth}
        \centering
        \includegraphics[width=\columnwidth]{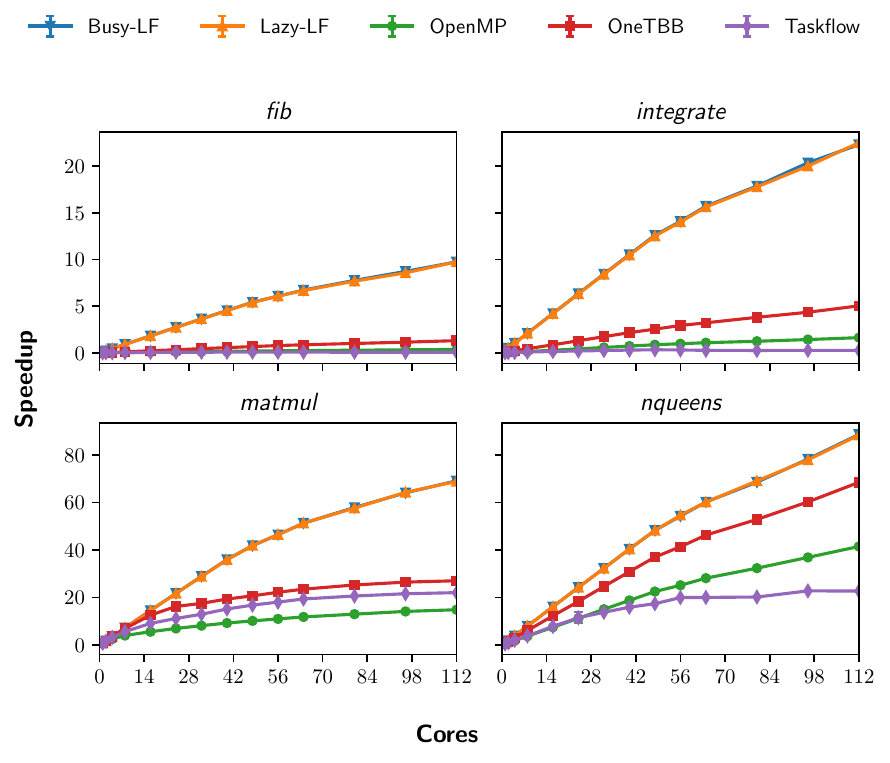}
        \caption{\label{fig::all}Parallel speedup -- \cref{eq::speedup}. }
    \end{subfigure}
    \hfill
    \begin{subfigure}[b]{0.49\textwidth}
        \centering
        \includegraphics[width=\columnwidth]{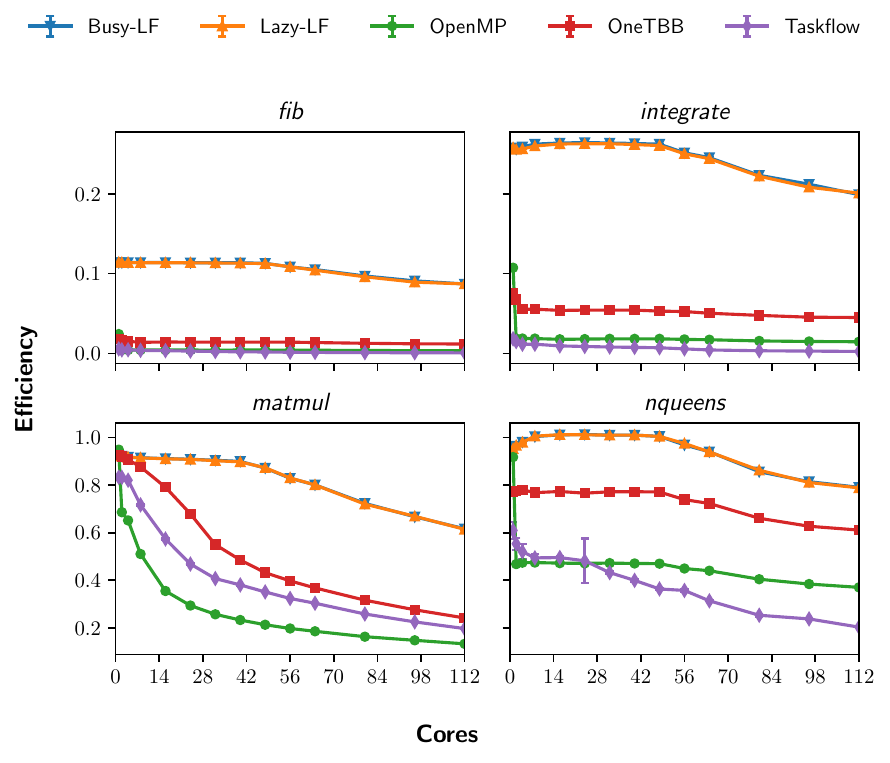}
        \caption{\label{fig::all_r}Parallel efficiency -- \cref{eq::eff}.}
    \end{subfigure}
    \vspace{0.5\baselineskip}
    \caption{Classical benchmarks, here \emph{Busy-LF} and \emph{Lazy-LF} refer to libfork's busy and lazy schedulers respectively.}
    \label{fig::trad}
\end{figure*}

\begin{figure*}[t]
    \centering
    \begin{subfigure}[b]{0.49\textwidth}
        \centering
        \includegraphics[width=\columnwidth]{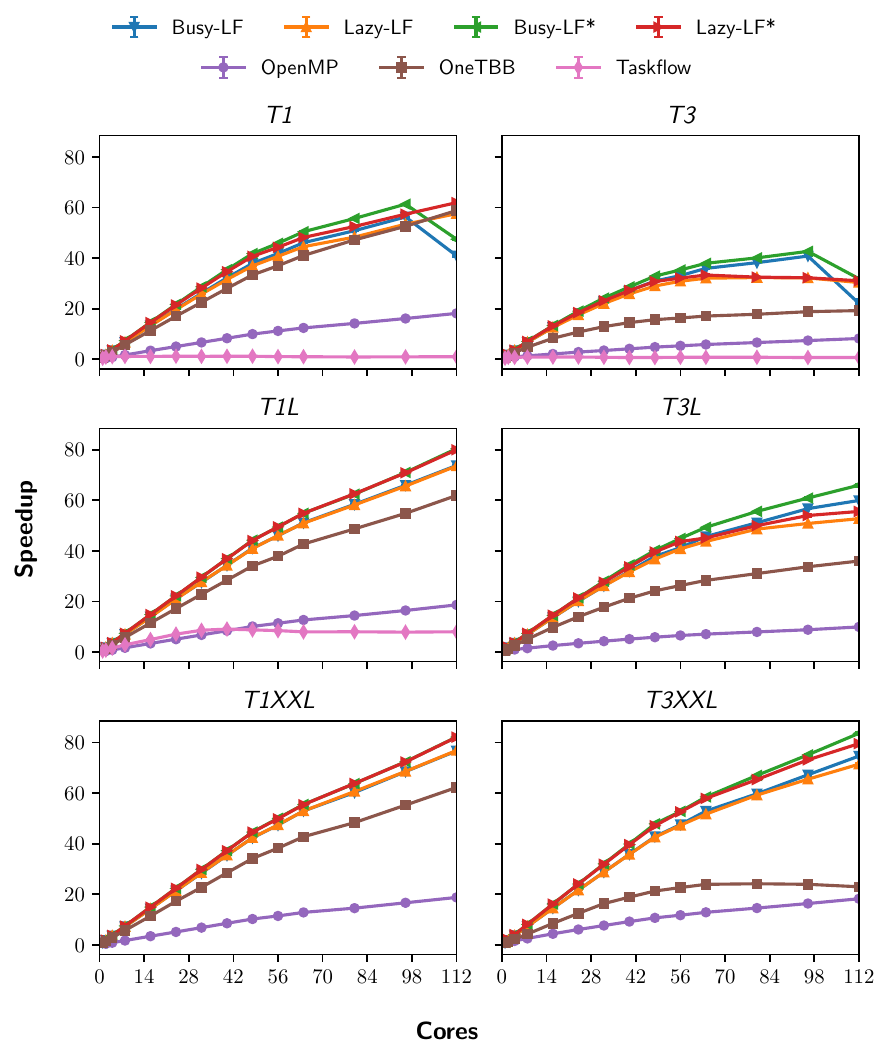}
        \caption{\label{fig::uts}Parallel speedup -- \cref{eq::speedup}.}
    \end{subfigure}
    \hfill
    \begin{subfigure}[b]{0.49\textwidth}
        \centering
        \includegraphics[width=\columnwidth]{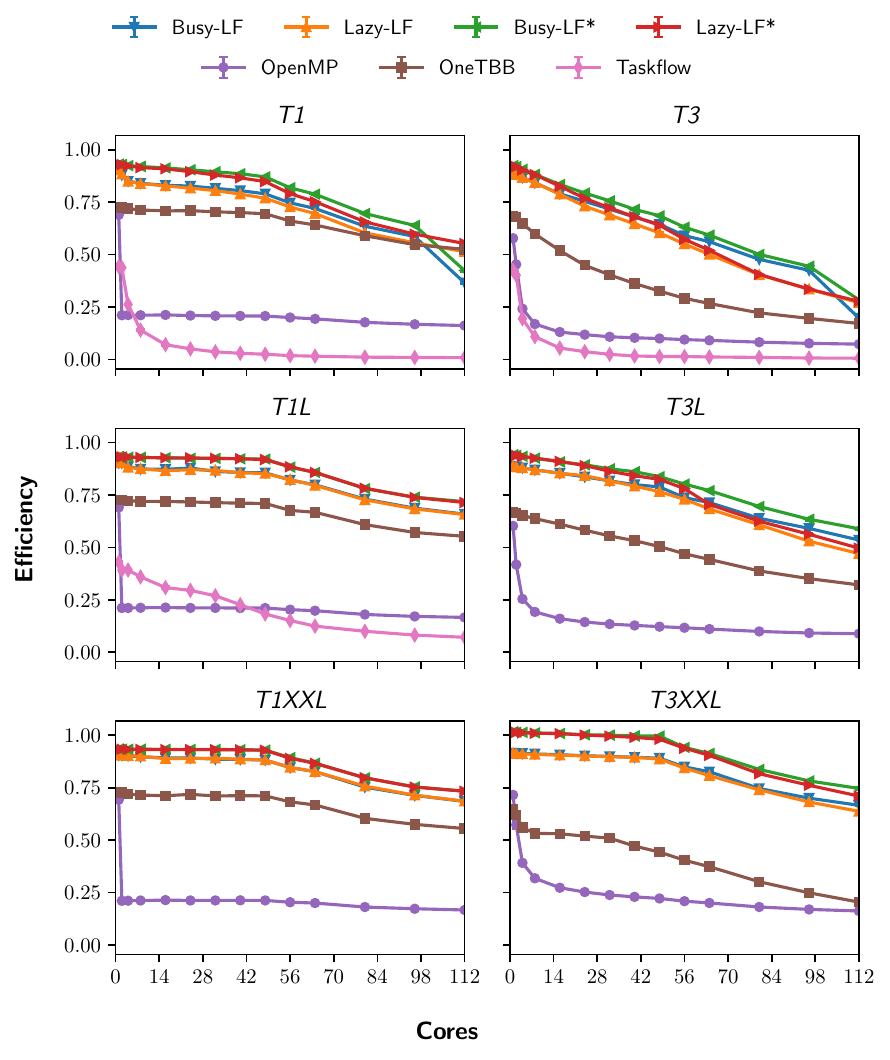}
        \caption{\label{fig::uts_r}Parallel efficiency -- \cref{eq::eff}.}
    \end{subfigure}
    \vspace{0.5\baselineskip}
    \caption{UTS benchmarks, here \emph{Busy-LF} and \emph{Lazy-LF} refer to libfork's busy and lazy schedulers respectively. The libfork benchmarks marked with a `*' used a modified algorithm utilizing libfork's stack allocation API (see \cref{sec::stalloc}) instead of heap-allocating space for return values. The taskflow benchmarks for T3L, T1XXL and T3XXL exhausted the available memory (500 GiB) and could not run to completion.}
    \label{fig::new_school}
\end{figure*}

We evaluated libfork with two sets of benchmarks. The first is a representative subset of the classical FJ/work-stealing benchmarks~\cite{Frigo1998, Yang2016, Schmaus2021} and the second is the unbalanced tree search (UTS) benchmark family~\cite{Olivier2007-ty}. The benchmarking parameters are detailed in \cref{tab::bench}.

\begin{table}
    \centering

    \caption{\label{tab::bench}Summary of benchmark parameters. For the UTS benchmarks, a \emph{geometric} tree has $t = 1$ and $a = 3$ while a \emph{binomial} tree has $t = 0$ and $b = 2000$. Unless otherwise specified, all other UTS parameters are defaulted.}

    \small

    \begin{tabular*}{\columnwidth}{@{}l@{\extracolsep{\fill}}ll@{}}
        \toprule
        Name      &  Description           & Parameters                       \\
        \midrule
        fib       & Recursive Fibonacci    & $n = 42$                         \\
        integrate & Numerical integration  & $n = 10^4, \epsilon=10^{-9}$     \\
        matmul    & Matrix $\times$ Matrix & $n = 8192$                       \\
        nqueens   & N-queens problem       & $n = 14$                         \\
        T1        & Small geometric tree   & $d=10, b=4, r=19$                \\
        T1L       & Large geometric tree   & $d=13, b=4, r=29$                \\
        T1XXL     & Huge geometric tree    & $d=15, b=4, r=19$                \\
        T3        & Small binomial tree    & $q=0.124875, m=8, r=42$          \\
        T3L       & Large binomial tree    & $q=0.200014, m=5, r=7$           \\
        T3XXL     & Huge binomial tree     & $q=0.499995, m=2, r=316$         \\
        \bottomrule
    \end{tabular*}
\end{table}

We compared libfork to Intel's TBB~\cite{Kukanov2007} and taskflow~\cite{Huang2019} which -- like libfork -- are both pure, portable library-implementation of FJ parallelism. We also made comparisons with openMP~\cite{Ayguade2009} (specifically the libomp implementation from the LLVM project~\cite{Lattner}); as a custom compiler front-end, openMP could access global-optimizations not accessible to the other libraries hence, it is not fully comparable. However, as it has been integrated into the major \Cpp compilers (GCC, Clang, ICC and, MSVC), we decided to include it for reference.

\subsection{Benchmark methodology}

\begin{table*}[b]
    \centering

    \caption{\label{tab::mem}Fitted exponential, $n$, from \cref{eq::mem_fit} and the data in \cref{fig::mem}; errors are estimated from the fit's covariance matrix.}

    \small

    \begin{tabular*}{\textwidth}{@{}l@{\extracolsep{\fill}}lllll@{}}
        \toprule
        Benchmark & Lazy-LF & Busy-Lf & TBB & OpenMP & Taskflow  \\
        \midrule
        Recursive Fibonacci     & $0.86 \pm 0.08$ & $0.93 \pm 0.06$ & $1.06 \pm 0.03$ & $1.20 \pm 0.10$ & $0.00 \pm 0.03$ \\
        Numerical integration   & $0.96 \pm 0.10$ & $0.94 \pm 0.06$ & $1.04 \pm 0.03$ & $1.07 \pm 0.09$ & $0.00 \pm 0.03$ \\
        Matrix $\times$ Matrix  & $0.88 \pm 0.08$ & $1.09 \pm 0.11$ & $1.07 \pm 0.03$ & $1.04 \pm 0.13$ & $0.00 \pm 0.08$ \\
        N-queens problem        & $0.94 \pm 0.03$ & $1.05 \pm 0.07$ & $1.11 \pm 0.03$ & $1.30 \pm 0.07$ & $0.00 \pm 0.37$ \\
        T1                      & $1.06 \pm 0.05$ & $1.04 \pm 0.03$ & $0.78 \pm 0.02$ & $1.18 \pm 0.06$ & $0.46 \pm 0.14$ \\
        T1L                     & $1.08 \pm 0.05$ & $0.99 \pm 0.06$ & $0.93 \pm 0.06$ & $1.23 \pm 0.08$ & $0.99 \pm 0.02$ \\
        T1XXL                   & $0.86 \pm 0.10$ & $0.90 \pm 0.03$ & $0.92 \pm 0.03$ & $1.08 \pm 0.07$ & -             \\
        T3                      & $0.83 \pm 0.05$ & $0.87 \pm 0.05$ & $0.67 \pm 0.03$ & $0.90 \pm 0.03$ & $0.64 \pm 0.15$ \\
        T3L                     & $0.43 \pm 0.03$ & $0.52 \pm 0.03$ & $1.00 \pm 0.03$ & $1.01 \pm 0.01$ & -             \\
        T3XXL                   & $0.38 \pm 0.01$ & $0.36 \pm 0.01$ & $0.86 \pm 0.01$ & $1.01 \pm 0.01$ & -             \\
        \bottomrule
    \end{tabular*}
\end{table*}

All benchmarks were run on a NUMA machine with two Intel(R) Xeon(R) Platinum 8480+ CPUs at \num{2.00}~GHz with boost enabled up to \num{3.8}~GHz. Each CPU socket had its own NUMA node and \num{56} cores, totaling \num{112} cores. The total available RAM was 500 GiB split over both sockets. The benchmarking code was compiled using Clang version 18.0.0\footnote{Using commit \texttt{c4b795df8075b111fc14cb5409f7138c32313a9b}, from the source available at \url{https://github.com/llvm/llvm-project.git}} with the highest optimization setting and runtime asserts disabled. We used the libomp distribution bundled with Clang, libfork version 3.5.0\footnote{Using commit \texttt{8e4ab70ee1c8b200a3e89fd467b7acadbb7c6d9e}, from the source available at \url{https://github.com/ConorWilliams/libfork.git}}, taskflow version 3.6.0 and, TBB version 2021.10.0.

Timings were performed using the Google benchmark library. All benchmarks were repeated until a minimum time had elapsed. Measurements of the maximum resident set size (MRSS) during the execution of a benchmark were performed with the GNU \texttt{time} utility. These are quantized to 4 KiB increments. Unless otherwise stated, all benchmark were run \num{5} times. We report the median and standard deviation of these measurements.

\subsection{Results}

\subsubsection{Execution time} The results of the execution time for the benchmarks are presented in \cref{fig::trad} and \cref{fig::new_school}. The parallel speedup of a program is defined as:
\begin{align}
    \text{Speedup} = \frac{T_s}{T_p} \label{eq::speedup}
\end{align}
Following \cref{eq::time_ideal}, for a program with sufficient parallelism (\ie $T_1 \gg T_\infty$) the speedup is expected to be linear. Additionally, the parallel efficiency is defined as:
\begin{align}
    \text{Efficiency} = \frac{\text{Speedup}}{P} \label{eq::eff}
\end{align}
which, for a linear scaling framework, should approach one as $T_1$ approaches $T_s$. At $P = 1$, the efficiency is equal to $T_s / T_1$, which directly measures the overhead of a framework in the absence of  communication interference, \eg stealing, cache-thrashing, lock-contention, \etc.

\begin{figure}
    \centering
    \includegraphics[width=\columnwidth]{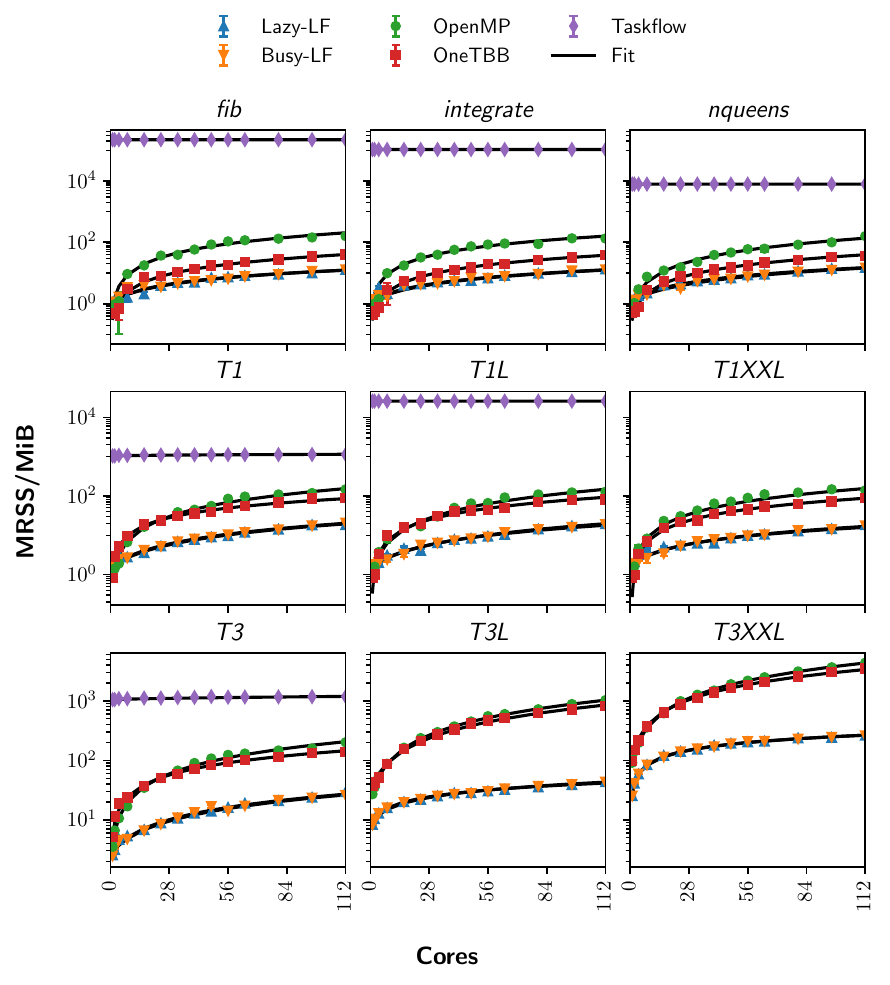}
    \caption{\label{fig::mem}Peak memory consumption during the benchmarks, fit to \cref{eq::mem_fit}. The taskflow benchmarks for T3L, T1XXL and T3XXL exhausted the available memory (500 GiB) and could not run to completion. The matrix multiplication benchmark has been excluded as the MRSS is dominated by the allocation of the input/output matrices.}
\end{figure}

\subsubsection{Memory consuption} The results of the memory consumption during the benchmarks are presented in \cref{fig::mem}. We fit the data to the power law:
\begin{align}
    \text{MRSS} \approx a + b M_1 P^n \label{eq::mem_fit}
\end{align}
with fitting parameters $a$, $b$ and $n$. The fitted exponents are presented in \cref{tab::mem}.

\subsection{Discussion}

All computations are fully-strict and libfork is a greedy scheduler that maintains the busy-leaves property~\cite{Blumofe_1999} hence, libfork's execution time should scale as \cref{eq::time_ideal}. This linear scaling is apparent in almost all the benchmarks up to around $56$ cores (half the total). After this point a second linear scaling continues but with a shallower gradient. This is probably due to the CPU clock-boost throttling back; as more of the cores become active the CPU is no-longer able to maintain the boosted clock speeds due to the additional thermal load.

\subsubsection{Classic benchmarks}

Across all the classic benchmarks, libfork's lazy and busy schedulers have almost identical performance. This is to be expected as the classic benchmarks spawn a large number of tasks. These tasks are easily distributed across the available cores. Their recursive structure means a thief always steals the largest task a victim has available, hence, a lazy worker is unlikely to ever sleep.

\paragraph{Fibonacci and integration} We see libfork significantly outperforms the other libraries in the Fibonacci and integration benchmarks. This is highlighted at $112$ cores where libfork is $7.5\times$ and $4.5\times$ faster then TBB and $24\times$ and $14\times$ faster than openMP. These are very fine-grained benchmarks that predominantly-test scheduling overhead. In particular, for Fibonacci (which has only a few instructions per-task), the overheads, $T_1 / T_S$, where \num{8.8}, \num{41}, \num{57} and, \num{180} for libfork, openMP, TBB, and, taskflow respectively. This cannot be due to libfork's WSQ implementation as the same one is used in taskflow~\cite{Lin2020}. Instead, the overhead of creating a libfork task is much lower; this could be dominated by heap allocations in the other libraries.

\paragraph{Matrix multiplication} The matrix multiplication benchmark computes the product of two matrices whose size was $\bigO{\text{GiB}}$ (much larger than the CPU caches) hence, cache locality and NUMA are important considerations. As the only continuation stealer, libfork scales much better than the other candidates. This is particularly apparent in \cref{fig::all_r}, where none of the other candidates manages a horizontal efficiency.

\paragraph{N-queens problem} The n-queens benchmark is perhaps the easiest to schedule as each task contains a substantial amount of work while the cache locality concerns are much less than the matrix multiplication benchmark. We see most of the frameworks manage linear scaling but, libfork still outperforms them, probably due to its reduced scheduling overhead.

\paragraph{Memory scaling }The memory profiles (\cref{fig::mem}) for the Fibonacci, integration and n-queens benchmarks all follow a similar pattern. Strikingly, taskflow consistently consumes \num{2}--{4} orders-of-magnitude more memory than libfork. This could be due to taskflows's internal caching of the (exponentially) many tasks produced by the recursive benchmarks. Furthermore, looking at the fitted exponents in \cref{tab::mem}, we see taskflow seems to allocate all these tasks regardless of the amount of parallelism. OpenMP and TBB perform better but still consume up to $12\times$ and $3.1\times$ more memory than libfork. This is probably due to a combination of libfork's segmented stacks reducing wasted memory and potentially a smaller metadata overhead per-task. The fitted exponents for libfork are all less than $1$, confirming that libfork's memory scaling is better than its theoretical bound. The sub-linear scaling is probably due to some stack space being shared between cores. Contrastingly, TBB consistently scales with an exponent just above $1$ and openMP has exponents as high as $1.3$. Some of this could be due to the \texttt{malloc} implementation fragmenting the task allocations but, this is likely a consequence of child-stealing. Libfork's coefficient $b$ in \cref{eq::mem_fit} never exceeds $0.2$, this is much less than the theoretical bound, $2c + 3 = 99$, of \cref{thm::memory}, which further reinforces the looseness of the bound.

\subsubsection{UTS benchmarks}

The geometric UTS benchmarks (\ie the T1 family) form a similar tree structure to the classic benchmarks; that is, the expected size of a subtree increases with proximity of the subtree's root to the root. The binomial UTS benchmark (\ie the T3 family) has irregular, self-similar subtrees and is ``an optimal adversary for load balancing strategies, since [\ldots] the expected work at all nodes is identical''~\cite{Olivier2007-ty}.

\paragraph{Small trees} The T1 and T3 benchmarks are relatively small benchmarks, taking only $10$ms and $13$ms to execute with libfork at $112$ cores. This explains libfork's sub-linear -- sometimes negative -- time scaling and low efficiency in \cref{fig::new_school}; cores run out of work and attempt spurious stealing of small tasks, which adds considerable overhead. Interestingly, the lazy scheduler mitigates this -- by sleeping the workers that cannot find work -- and does not exhibit the negative scaling. Notably, TBB handles this work-starved environment well.

\paragraph{Geometric trees} For the remainder of the T1 family, libfork demonstrates similar linear scaling as in the classical benchmarks because, the underlying program's DAGs have a similar structure. As the recursion depth is similar across the geometric trees, the memory consumption remains roughly constant for libfork, openMP and TBB. Again, taskflow continues to allocate enough memory for all the tasks ever-spawned, this eventually exhausts the system's memory for T1XXL and terminates the benchmark.

\paragraph{Binomial trees} Despite the increased difficulty in load balancing, libfork demonstrates linear time-scaling for the remainder of the binomial family. The T3L benchmark continues to trigger frequent steals, incurring overhead, and resulting in a lower efficiency. The binomial trees have a larger recursion depth hence, requiring more memory. As with the T1 family, taskflow exhausts the available memory and fails to complete the benchmarks. Libfork does a much better job at minimizing allocations; in the T3XXL benchmark, libfork requires $13\times$ less memory than TBB and $17\times$ less memory than openMP, despite delivering at least a $3.6\times$ speedup over both. In \cref{tab::mem}, we see the fitted exponents dropping to much less than one. This is because the depth of each leaf node in the DAG is binomially-distributed hence, the average worker will be using much less than the maximum stack space.

\paragraph{Stack allocation API} The measurements annotated with a `*' in \cref{fig::new_school} used a modified algorithm, utilizing libfork's stack allocation API (see \cref{sec::stalloc}). This brings a small performance enhancement in all the UTS benchmarks, as it reduces the number of heap allocations and increases cache locality.

\section{\label{sec::conc}Conclusions}

We have shown how the operations of continuation-stealing fork-join parallelism can be mapped to the primitives of stackless coroutines and developed libfork: a lock-free, wait-free fine grained, NUMA aware, fully decentralized (no global queues), weak-memory-model optimized, parallelism library. By utilizing \Cppn{20}'s coroutines libfork is, to the best of the author's knowledge, the only fully-portable continuation-stealing \Cpp tasking library. Libfork achieves linear time-scaling, performing up to: $7.5\times$ faster than Intel's TBB, $24\times$ faster than openMP (libomp) and, $100\times$ faster than taskflow. Similarly, libfork achieves linear memory-scaling, through its integration of user-space segmented stacks to store coroutine frames, consuming up to: $19\times$ less memory than Intel's TBB, $24\times$ less memory than openMP (libomp) and, several orders-of-magnitude less memory than taskflow.

Libfork is an ongoing development; coroutines are a relatively new addition to the \Cpp language. Hopefully, future language additions, such as asynchronous resource-acquisition-is-initialization (RAII), will allow for usability and performance enhancements. Future compiler optimizations, particularly work on the heap-allocation-elision-optimization (HALO) which could completely elide allocations for non-recursive coroutines, may bring library-based solutions on-par with language level solutions. For now, libfork enables programmers to expose finer-grained parallelism than previously possible, a necessary step to fully utilizing available hardware.

\section*{Artifacts}

The source code for libfork and its benchmark suite are available online: \url{https://github.com/ConorWilliams/libfork}

\section*{Thanks}

We would like to thank Patrick R.L. Welche for reviewing the manuscript and providing stimulating discussions.

\section*{Acknowledgements}

We gratefully acknowledge the funding received from the EPSRC via the CDT in Computational Methods for Materials Science (Grant number EP/L015552/1). We also acknowledge Rolls-Royce Plc for the provision of funding. All information and foreground intellectual property generated by this research work is the property of Rolls-Royce Plc. 
\renewcommand*{\bibfont}{\footnotesize}

\printbibliography

\end{document}